# Ultrafast Photocurrent Hysteresis in Photoferroelectric *α*-In$_2$Se$_3$

Zhen Lei[1], Jiawei Chang[1], Qiyi Zhao[2], Jian Zhou[3], Yuanyuan Huang[1]*, Qihua Xiong[4]*, and Xinlong Xu[1]*

[1] Shaanxi Joint Lab of Graphene, State Key Lab Incubation Base of Photoelectric Technology and Functional Materials, International Collaborative Center on Photoelectric Technology and Nano Functional Materials, Institute of Photonics & Photon-Technology, Northwest University, Xi'an 710069, P. R. China
E-mail: yyhuang@nwu.edu.cn, xlxuphy@nwu.edu.cn

[2] School of Science, Xi'an University of Posts & Telecommunications, Xi'an 710121, China

[3] Center for Alloy Innovation and Design, State Key Laboratory for Mechanical Behavior of Materials, Xi'an Jiaotong University, Xi'an 710049, China

[4] State Key Laboratory of Low-Dimensional Quantum Physics and Department of Physics, Tsinghua University, Beijing 100084, P.R. China
E-mail: qihua@ntu.edu.sg

*Abstract:* The photon-electron interactions are generally volatile and the intricate multiphysics details of photoexcited carrier dynamics are not yet distinguished. How to nonvolatile control the physical state through all-optical means and clarify the intricate physical processes has been a long-term goal pursued in polar materials. Photoferroelectric *α*-In$_2$Se$_3$ holds the great potential for capturing multimodal nonvolatile states due to the spontaneous reversible in-plane and out-of-plane polarizations and its tunable light-matter interactions arising from the electronic degree of freedom. Here we uncover a nonvolatile zero-bias ultrafast photocurrent hysteresis response with an all-optical scheme, diagnosed by in-plane and out-of-plane terahertz waves emitted from the photoferroelectric *α*-In$_2$Se$_3$. The mechanism of such ultrafast photocurrent hysteresis emerges as a result of anomalous bulk linear and circular photovoltaic effect synchronously driven by local polarization rearrangement. Utilizing anisotropic ferroelectric kinetics-induced relative phase between the in-plane and out-of-plane directions, we further show flexibly selective chirality, tunable rotational angle, and optimizable ellipticity of terahertz wave polarizations. Our finding offers a promising avenue towards direct ultrafast nonvolatile processing of photocurrent signals through an all-optical scheme.

**Introduction**

Nonvolatile control over the physical properties of quantum materials through ultrafast light-matter interaction has been a long-term objective in the all-optical manipulation field[1-3]. Photoferroelectric materials, combing both polarization nonvolatility and rich photophysical properties arising from the delicate coupling between light polarization and electronic symmetry, are anticipated to serve for multimodal nonvolatile memory and optoelectronic devices[4-6]. Of particular interest is that polarization reversal is interlocked with the unique properties of spin-orbit coupling (SOC) and real-space electron shifting in materials[7-10]. These characters indicate that the ferroelectric polarization reversal is usually associated with tunable spin and charge ultrafast behavior, and vice versa. Unfortunately, most currently used manipulation techniques of switching ferroelectric polarization require the implementation of electrodes and wires contacting and patterns for applying voltages[11-19]. This makes the system response time is typically on the order of a few nanoseconds or even longer. It would be much more efficient for ultrafast read/write operations of spin, charge, and other collective electronic responses, if one can shrink the typical timescale within a few picoseconds[19-21].

Progress in the combination of photoferroelectrics and femtosecond (fs) laser technology enables one to overcome this difficulty[19]. Near-infrared pulsed light is capable of providing an all-optical poling bias of over one trillion times per second to apply fs duration electric fields to photoferroelectrics. Thereby correlated ultrafast behavior of photogenerated carriers could unravel the ultrafast evolution dynamics under polarization switching with sub-terahertz temporal resolutions[22]. The peculiar properties of anomalous bulk photovoltaic (BPV) effect is triggered burst point for large open-circuit voltage surpassing the bandgap limitation and switchable photocurrents engendered by light polarization[23-24]. It also reflects the geometric feature of electronic wavefunction, giving rise to peculiar topological features in semiconductors. Owing to the symmetry arguments, the BPV effect is closely connected with the ferroelectric domain patterns and the dynamics of domain walls[21]. Strikingly, the switching of the light-induced zero-bias BPV effect in all directions typically coincides with the spontaneous polarization[25]. Therefore, the anomalous BPV effect could be perceived not only as a visualized probe but as a fascinating control knob for manipulating the nonvolatile on/off state of the switchable ferroelectric polarization[26-27]. However, owing to the complicated phase-transition kinetics and complex interactions among light, lattice, and carriers, exploring a directly visualized diagnosing technology and corresponding material platform remains a huge challenge. This greatly hinders our understanding of how ferroelectric ordering couples to the transport of electronic degrees of freedom and the utility of opto-mechano-electric devices.

Here we present a straightforward yet highly efficient method to examine the nonvolatile BPV effect and uncover the peculiar interaction of photoexcited nonequilibrium carriers and ferroelectric order in photoferroelectrics $\alpha$-$In_2Se_3$. According to the instantaneous snapshot of symmetry, phase, and polarization diagnosed by in-plane and out-of-plane time-resolved terahertz (THz) waves, we analyze the mechanism of ferroelectric order modulated light-current conversion according to the instantaneous symmetry variation.

## Results and Discussion

**Multiphysics THz approach of in-plane and out-of-plane photocurrents.** The $\alpha$-In$_2$Se$_3$ sample has a rhombohedral (3$R$) structure with $R3m$ space group (characterizations in Supplementary Note 1). The side-view (I, Fig. 1a) shows a quintuple layer (QL) of $\alpha$-In$_2$Se$_3$, stacking in a sequence of Se-In-Se-In-Se. The top-view (II, Fig. 1a) gives a typically hexagonal lattice structure with Se or In atoms arranged in a trigonal lattice. In particular, as shown in Fig. 1a (III), the shifting of central Se atoms in a QL creates switchable in-plane ($P_{IP}$) and out-of-plane ($P_{OOP}$) spontaneous reversible ferroelectric polarization[28].

Figure 1b depicts a reflection-type schematic for generating THz pulses in the ferroelectric $\alpha$-In$_2$Se$_3$. A near-infrared 800 nm pulsed light excites the $\alpha$-In$_2$Se$_3$ sample and launches the in-plane [$J_y(t)$] and out-of-plane [$J_x(t)$ and $J_z(t)$] transient pulsed photocurrents. Following, the transient photocurrents instantly radiate time-resolved THz fields ($\boldsymbol{E}(t)=E_{xz}(t)\boldsymbol{p}+E_y(t)\boldsymbol{s}$ with the coordinate details in Supplementary Note 2), which consist of $p$-polarized $E_{xz}(t)$ in the $x$-$z$ plane (related to $J_{xz}(t)$) and the $s$-polarized $E_y(t)$ (related to $J_y(t)$) along $y$-axis, respectively.

**Nonvolatile ultrafast photocurrents based on LBPV effect.** To determine the potential physical mechanisms, we analyze the effect of the pump fluence on THz wave emission. As detailed in Supplementary Note 3, both $E_{xz}$ and $E_y$ grow linearly with the pump fluence ($E_{\text{THz}} \propto I$) (Fig. S3). The linear behavior confirms the leading role of second-order nonlinear response in the THz emission process, rather than higher-order[29] or saturation-type[30-31] photocurrents. Furthermore, we rule out photon drag effect due to the requirement for high carrier mobility and carrier concentration[32]. Consequently, the linearly light induced BPV effect (LBPV, or shift current $J_{\text{shift}}$ for time reversal invariant materials) is the primary mechanism to the THz emission process.

To explore the features of ultrafast photocurrents in $\alpha$-In$_2$Se$_3$, we have measured the dependence of $E_{xz}(t)$ and $E_y(t)$ as a function of light polarization angle ($\zeta$) by adjusting a $\lambda$/2-waveplate. The angle-dependence over a period (0°~360°) of $E_{xz}(t, \zeta)$ peak-valley values are shown in Fig. 2a. Clearly, $E_{xz}$ depends strongly on the light polarization and the contour mapping of $E_{xz}(t, \zeta)$ further confirms the $2\zeta$-dependent tendency. This tendency can be usually fitted with the tensor analysis-based LBPV model[11]. However, contrary to early observations[11, 33], the $E_{xz}(t, \zeta)$ results match inexactly with the predicted optical nonlinear scenario (details in Methods and Supplementary Note 4), especially at 90° and 270°. This implies that the physical procedure is not pure LBPV, but is mixed with peculiar light-matter interactions with specific ferroelectric order. In addition, the upward offset of the $E_{xz}$ trend (Fig. 2a) is caused by the $\zeta$-independent term related to the nonzero nonlinear conductivity coefficients $\sigma_{xzx}$, $\sigma_{zxx}$, and $\sigma_{zzz}$ (details in Supplementary Note 4).

By contrast, $E_y$ strikingly enhances at the $A_1$ and $A_2$ polarization angles, and one clearly observes that the contour mapping of $E_y(t, \zeta)$ THz fields presents anomalous fourfold anisotropic $\zeta$-dependency (Fig. 2b). These results differ significantly from the radiation properties of THz waves with linearly-polarized light in previous reports[33]. This anomalous dependency cannot be well fitted by the LBPV results for a single structural phase (details in Methods and Supplementary Note 4). This implies that the THz wave generation here involves multiple coupling physical processes.

Intriguingly, when the half waveplate reaches to 90° and 270° (Fig. 2b) with the same s-polarized light excitation, we observe a remarkable delay in the waveform of $E_{xz}(t)$ THz electrical fields (Fig. 2c). Moreover, the polarity of the $E_y(t)$ is switched oppositely by the same s-polarized light excitation (Fig. 2c). These anomalous light-induced delay/switchable phenomena appear to be linked with non-volatility triggered by the inherent $P_{IP}$ and $P_{OOP}$ in ferroelectric α-In$_2$Se$_3$. This delay and switchable phenomenon has never been observed in traditional semiconductors such as p-InAs (details in Supplementary Note 5), WS$_2$[31], perovskite CsPbBr$_3$[29], and Weyl semimetal TaAs[33], which further suggests the unique nature in ferroelectric α-In$_2$Se$_3$.

To determine this anomaly, it is necessary to further explore the potential photophysical mechanism in α-In$_2$Se$_3$. Apart from the time-domain information, phase can provide further insight into the complex photophysical phenomena[33-34]. As shown in Fig. 2d, we display the normalized electromagnetic waves at the central frequency (~1.2 THz) of $E_{xz}(t)$ and $E_y(t)$ for the same s-polarized light. We note that the $E_{xz}(t)$ is

0.2π-phase delayed, while the $E_y(t)$ shows a π-phase delay, signifying a phenomenological photoferroelectric behavior of zero-bias nonlinear photocurrents. To quantify this polarization-dependent behavior, we show the tendency of $E_{xz}$ and $E_y$ in an cyclical setting of ζ-values (0°→90°→180°→90°→0°), which exhibits evidently a photoferroelectric hysteresis pattern (Fig. 2e,f). This $J_\lambda^L$-$E_\zeta$ (λ=xz, y) pattern is analogous to the common static ferroelectric polarization (see Supplementary Fig. S1) with obvious multivalued $J_\lambda^L$ (λ=xz, y) at the same light polarization.

To understand this light-polarization dependent hysteresis, the light-polarization dependence of both shift current and all-optical poling to α-In$_2$Se$_3$ should be considered simultaneously, to capture the general feature of the photocurrent hysteresis loop. Primarily, as detailed in Methods, there exists a readily apparent correlation between shift current $J_{shift}$ and polarization $\mathcal{P}$, as they are both rooted with geometric phase of electronic wavefunctions[35-36]. The $J_{shift}$ can be estimated as (details in Supplementary Note 6)[37]

$$\boldsymbol{J}_{shift} \simeq \sigma^{(1)}(e/\hbar\omega)|E_\xi|^2 \boldsymbol{R} \tag{1}$$

where $\sigma^{(1)}$ is the linear photoconductivity and $\boldsymbol{R}$ is the shift vector. It not only reveals the second-order response in the light field $E_\zeta$, but also involves the quantum origin ($\boldsymbol{R}$). Notably, one observes that the $\sigma^{(1)}$ present an inherent electrical hysteresis as the $E$-$P$ curve in ferroelectric materials. Furthermore, the light field $E_\zeta$ can also be perceived as an applied fs-duration electric field, and provides an all-optical poling to α-In$_2$Se$_3$ that is sensitively related to the light polarization. Therefore, the output $J_{shift}$ exhibits a hysteresis for a cyclical loop of the light polarization as shown in Fig. 2e,f, with different values along clockwise path (adc) compared with the anticlockwise path (cba). The $E_{xz}$ is mainly related to the out-of-plane LBPV with a small hysteresis (Fig. 2e), while the $E_y$ is mainly related to the in-plane LBPV, which demonstrate the flipping of ferroelectric order (Fig. 2f). The difference between the in-plane and the out-of-plane hysteresis is subject to non-synchronous polarization switching in ferroelectrics with different symmetry transformation rules[38]. In addition, the all-optical poling to α-In$_2$Se$_3$ leads to a sinusoidal dependence of $E_{xz}$ and $E_y$ on the azimuthal angle (see Supplementary Note 7), which is completely different from the triple-rotational symmetry dependence of non-ferroelectric materials in the 3$m$ point group[39].

Next, we phenomenally analyze the hysteresis triggered by the light-polarization in α-In$_2$Se$_3$. The $E_{xz}(t)$ and $E_y(t)$ is strongly related to the ferroelectric polarization $P_{IP}$ and $P_{OOP}$ which would be ultrafast modulated by the nonequilibrium carriers excited by the fs laser[40]. Namely, the ultrafast motion of photoexcited carriers reduces the original screening field $E_s$, leading to a net electric field Δ$E$ that is anti-parallel to the direction of $P_{IP}$ and $P_{OOP}$ (details in Supplementary Note 8). This internal field alteration in turn causes the $P_{IP}$ and $P_{OOP}$ to rearrange according to the incident light polarization[41]. Synchronously, the photoexcited nonequilibrium carriers would move toward the rearranged directions of $P_{IP}$ and $P_{OOP}$, accompanying with the phenomenological ferroelectric behavior. In the extreme case, when the total of Δ$E$ and $E_\zeta$ exceeds the coercive field, polarization reversal would be triggered as shown in the positions b and d in Fig. 2f. However, the weaker remanent polarization along the out-of-plane direction hinders the occurrence of polarization reversal, as evidenced by positions b and d in Fig. 2e.

**Nonvolatile ultrafast photocurrents based on CBPV effect.** In the following, we reveal the subtle interactions between helicity-dependent ultrafast circular light induced BPV photocurrents (denoted as CBPV effect) and ferroelectric orders. The CBPV effect scales as the velocity difference at $k$ and the imaginary part of quantum geometric metric tensor of electronic wavefunctions, namely, Berry curvature Ω($k$). Note that since Berry curvature transforms as a pseudovector, under time reversal symmetry it satisfies Ω($k$) = –Ω(–$k$). The Dresselhaus and Rashba type SOC breaks the velocity difference between the valence and conduction bands, at $k$ and –$k$. Remarkably, the electronic bands of α-In$_2$Se$_3$ undergoes Rashba-type spin-splitting (Extended data Fig. 1a) due to the presence of heavy-atoms In and Se, which satisfies the prerequisites of CBPV effect[28, 42]. As sketched in Fig. 3a, the nonequilibrium occupation of spin-polarized electrons yield time reversal broken-related photocurrent (i.e., injection current) under the circularly-polarized light excitation in *k*-space[13, 43] due to the angular momentum conservation and optical selection rule.

In Rashba-type spin-splitting systems[44-45], the CBPV current with opposite light helicities typically flows along opposite paths, resulting in the polarity inversion of THz wave (Fig. 3b). As shown in Fig. 3c, the polarity of $E_y(t)$ (in-plane component along *y*-axis) is reversed and the amplitude is almost equal when the light chirality changes from right- ($\delta^+$) to left-handed ($\delta^-$) polarized light (Fig. 3c). Such result verifies that the injection current is the predominant factor in generating $E_y(t)$ THz field due to the Rashba spin-orbit coupling in $\alpha$-In$_2$Se$_3$. In contrast to $E_y(t)$, the polarity of $E_{xz}(t)$ (out-of-plane component along *x-z* plane) remains for $\delta^+$ and $\delta^-$ excitation (Extended data Fig. 1b). This is because the $E_{xz}(t)$ THz field is generated from the two perpendicular photocurrents ($J_x(t)$ and $J_z(t)$), involving multiple photophysical processes.

To further clarify the in-plane and out-of-plane photocurrents, we turn to the dependence of $E_{xz}$ and $E_y$ on the pump fluence. In Fig. 3d, $E_y$ shows a reverse saturation trend as the pump fluence of $\delta^+$ and $\delta^-$ increases. This trend is the result of both energy and spin relaxation in helicity-dependent photoexcitation, leading to spin-polarized photocurrent bleaching[46], instead of the saturation from built-in electric field[31] or photothermoelectric[30] effect. However, due to the multiple photophysical processes, $E_{xz}$ increases (and saturates at the extreme) with the pump fluence for $\delta^+$ excitation, while increases linearly with that of $\delta^-$ (Extended data Fig. 1c). For $\delta^+$ excitation, we note that $E_{xz}$ has a weaker saturation effect than $E_y$, which would result from the combination of linear-dependent shift current and saturation-dependent spin-polarized current on the pump fluence. For $\delta^-$ excitation, the linear-dependence of $E_{xz}$ on pump fluence suggests shift current could dominate the emission of $E_{xz}(t)$.

To gain a deeper understanding of helicity-dependent photocurrents, we measure the $E_{xz}(t)$ and $E_y(t)$ THz fields as a function of orientation angle ($\psi$) of $\lambda/4$ waveplate (QWP). As shown in Fig. 4a, $E_{xz}(t, \psi)$ contour-mapping manifests modulation between $\delta^+$ and $\delta^-$ with asymmetric peak offset, which suggests obvious circular dichroism This result can be usually fitted by the second order CBPV components (see Methods and Supplementary Note 4). However, different from previous works[17, 33, 45-46], the $E_{xz}(t, \psi)$ results display a definite delay within 90°~180° compared with the CBPV theoretical results (red curve). This suggests that the photophysical process is impacted by a unique light-matter interaction that involves a specific ferroelectric order.

In comparison, $E_y(t, \psi)$ contour-mapping clearly reveals the polarity inversion of THz signal at $\delta^+$ and $\delta^-$ (Fig. 4b), reflecting stronger circular dichroism than $E_{xz}(t, \psi)$. Intriguingly, the maximal value of $E_y$ appears at the peculiar angles of 60° and 150°, which is different from the CBPV response as reported in $MoS_2$[46], organic-perovskites[45], and BiTeBr[44]. Note that the maximal values in these non-photoferroelectric materials generally appear at $\psi=\pm\pi/4$[44-46], whereas our results appear at a position with ~15° delay. Moreover, compared to the CBPV theoretical results (Fig. 4b), $E_y$ shows a significant delay over a period (0°~180°) with the light polarization. This further implies that THz wave emission from $\alpha$-$In_2Se_3$ involves multiple photophysical processes synergistically.

The tensorial theoretical analysis can give an intuitive view of the contribution of CBPV and LBPV under the elliptical light excitation. Figure 4c shows four fitting parameters $C_\lambda$, $L_{1\lambda}$, $L_{2\lambda}$, and $D_\lambda$ ($\lambda=y$ or $xz$) for the individual contributions of CBPV and LBPV effect, respectively. Here, $C_\lambda$ stands for the CBPV-induced $J_\lambda^C$; both $L_{1\lambda}$ and $L_{2\lambda}$ contribute to the LBPV response (or shift current $J_\lambda^L$); $D_\lambda$ represents the $\psi$-independent constant term, referring to the nonzero nonlinear coefficients. The magnitude order of coefficients clearly reveals that the contribution along the $x$-$z$ plane: $C_{xz} \approx L_{1xz} > L_{2xz}$ and $y$-axis: $C_y \gg L_{1y}, L_{2y}$. This implies unambiguous contribution of $C_y$ term to CBPV leads to the THz signal along the $y$-axis, and both LBPV and CBPV effects simultaneously dominate the process within $x$-$z$ plane.

To elucidate the delay features, we extract the phase information of the $E_{xz}(t, \psi)$ and $E_y(t, \psi)$ THz waveform at the central frequency (1.2 THz). Intriguingly, Fig. 4d shows that the phase of both $E_{xz}(t, \psi)$ and $E_y(t, \psi)$ display hysteresis features with the cyclical setting of $\psi$-values (0°→45°→0°→-45°→0°). Under $\delta^-$ and $\delta^+$ excitation, $E_{xz}(t)$ phases are approximately $0.4\pi$ and $-1.2\pi$ respectively, resulting in a phase difference of $-1.6\pi$. For this reason, the THz waveform polarity of $E_{xz}(t)$ is not flipped (Fig. 4a). On the other hand, $E_y(t, \psi)$ phases are approximately $-0.4\pi$ and $-1.4\pi$ (phase difference of $\pi$) under $\delta^-$ and $\delta^+$ excitation, which accounts for the polarity reversal of $E_y(t)$ (Fig. 4b). Apart from the phase feature, the peak-valley value of $E_{xz}$ and $E_y$ clearly form a polarization-dependent photoferroelectric hysteresis loop as shown in Fig. 4e,f. These typical hysteresis loops $J^C_{xz}$-$E_\psi$ (Fig. 4e) and $J^C_y$-$E_\psi$ (Fig. 4f) demonstrate the spin-related photoferroelectrics physics, recording the ferroelectric-like bistable states of spin-polarized photocurrents. This $\psi$-dependent hysteresis is also related to the helicity-dependent photocurrent and all-optical poling to $\alpha$-In$_2$Se$_3$ synchronically.

As detailed in Methods, the injection current ($J_{\text{inject}}$) is related to the quantum geometric phase from the Bloch wavefunction, which would be influenced by the photoferroelectric polarization with a hysteresis as discussed in the LBPV part. The $E_{xz}$ corresponds mainly to the out-of-plane multiple photocurrents with a butterfly hysteresis (Fig. 4e), while the $E_y$ is primarily associated with the in-plane injection current with a flipping of ferroelectric order hysteresis (Fig. 4f). This difference in the in-plane and out-of-plane hysteresis response reflects the different polarization switching with different symmetry transformation rules in $\alpha$-In$_2$Se$_3$[38]. In addition, these hysteresis loops are almost frequency-independent, which can acquire the same photoferroelectric physical parameters in a broad bandwidth with ultrafast response (details in Supplementary Note 10). Consequently, this nonvolatile Rashba physics coupled with ferroelectric polarization in ferroelectric $\alpha$-In$_2$Se$_3$ can be used as building block "1" and "0" states for the light logic circuit or a memory cell.

**Manipulation of THz wave polarization**. Note that a relative phase $\varphi_d$ is generated between $E_{xz}(t)$ and $E_y(t)$ as a result of the anisotropic photoferroelectric polarization kinetics along $P_{\text{OOP}}$ and $P_{\text{IP}}$ directions. This $\varphi_d$ is a prerequisite for THz polarization control, which can be controlled by tuning the angle $\zeta$ and $\psi$. In general, when $\varphi_d$ changes, both rotational angle $\beta^{\text{THz}}$ and ellipticity $\eta^{\text{THz}}$ typically change

(details in Supplementary Note 11). When $\zeta$ changes from 0 to $2\pi$ (Fig. 5a), we find $\varphi_d$ is almost 0 and $\pi$, which determines linearly-polarized states of far-field THz field as $\boldsymbol{E}(t)=E_{xz}(t)\boldsymbol{s}+E_y(t)\boldsymbol{p}$. As $\zeta$ varies from 0 to $\pi$ (Fig. 5b) and from $\pi$ to $2\pi$ (Fig. 5c), the polarization trajectories of THz fields rotate in the counterclockwise way. Moreover, the calculated ellipticity $\eta^{THz}$ of $\boldsymbol{E}(t)$ is almost 0.1, and the chirality of $\boldsymbol{E}(t)$ maintains right-handed.

To achieve further control of chirality and ellipticity, we manipulate the light helicity from $\delta^+$ to $\delta^-$ by changing the angle $\psi$. As illustrated in Fig. 5d, the THz field ($\boldsymbol{E}(t)$) exhibits an elliptically-polarized trajectory with right-handed chirality when excited with $\delta^+$ light. This result is consistent with the density mapping of $\boldsymbol{E}(t)$ polarization trajectory at the central frequency (1.2 THz). Accordingly, the calculated $\beta^{THz}$ is ~$5\pi/12$ rad, and $\eta^{THz}$ is ~0.36 for $\delta^+$ light excitation in Fig. 5d. In comparison, the polarization trajectory for $\delta^-$ light excitation (Fig. 5e) displays a left-handed chirality with $\beta^{THz}$ of ~$7\pi/12$ rad and $\eta^{THz}$ of ~0.23, which is confirmed by the trajectory density mapping of $\boldsymbol{E}(t)$ at 1.2 THz.

To gain a further insight into the THz polarizations, we manipulate the light polarization from linear to circular by rotating the $\psi$-values (details in Supplementary Note 12). As $\psi$ changes, the THz polarizations can be continuously manipulated from right-handed to left-handed. Moreover, the ellipticity of the $\boldsymbol{E}(t)$ could increase to $\eta^{THz}=0.6$ (Fig. 5f), and the variation period of $\beta^{THz}$ is approximately $\pi$ (Fig. S12). Compared to two-color laser scheme[47] and engineered materials as W/CoFeB/Pt heterostructure[48], we unexpectedly achieve the manipulation of greater ellipticity of $\boldsymbol{E}(t)$ and greater flexibility of chirality and rotational angle by coherently controlling the intrinsic $P_{IP}$ and $P_{OOP}$ properties of $\alpha$-In$_2$Se$_3$.

**Conclusions**

In summary, we have demonstrated a striking interlock among light, electron, and ferroelectric order in the photoferroelectric $\alpha$-In$_2$Se$_3$ without electrical connection. Employing THz spectroscopy as a powerful probe for in-plane and out-of-plane photocurrents, we have synchronically visualized the BPV effect involving specific ferroelectric order degree of freedom. We unexpectedly discover that the ultrafast photocurrent displays a photoferroelectric hysteresis in both in-plane and out-of-plane directions, triggered by the ultrafast pump light polarization. This allows us to control over the nonvolatile photophysical in a coherent and efficient manner, which is of much fundamental and applied physics interest. Furthermore, we flexibly control the chirality, rotational angle, and ellipticity of THz polarizations by manipulating the light polarization. This work realizes an all-optical control of the on/off behavior of light-induced ferroelectric hysteresis, which opens up new potential application platforms of ferroelectric materials.

# Figures

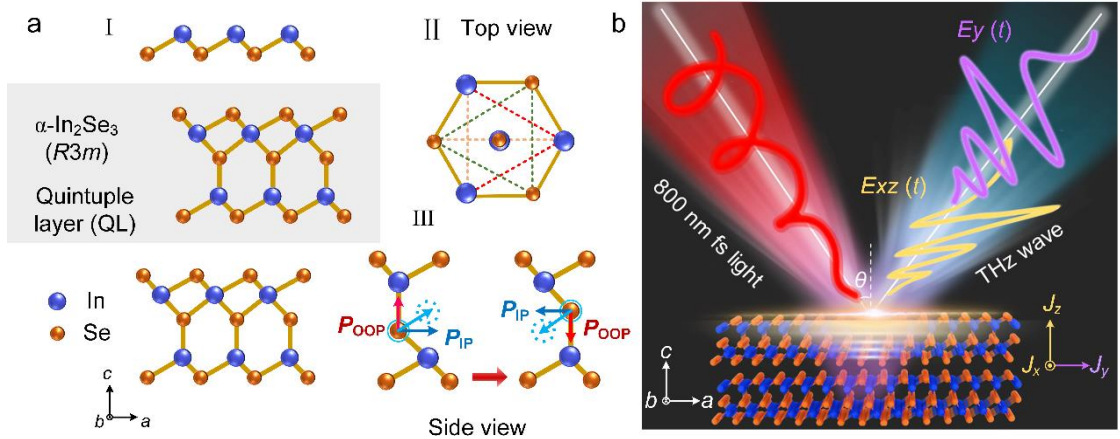

**Figure 1. Schematic of THz wave generation from 3R α-In₂Se₃ in a reflection configuration.** (a) (I) Three-dimensional atomic structure of 3R α-In$_2$Se$_3$ with Se atoms (orange) and In atoms (blue) following to the repeated Se-In-Se-In-Se arrangement, in which the gray area indicates a quintuple layer (QL). (II) Top-view of the single layer α-In$_2$Se$_3$ indicates the basically hexagonal atomic structure. (III) Side-view atomic structure of 3R α-In$_2$Se$_3$ indicting the pronounced in-plane (P$_{IP}$) and out-of-plane (P$_{OOP}$) polarization. The polarization direction of coherent P$_{IP}$ and P$_{OOP}$ can be switched by applied ultrafast light field. (b) Schematic illustration of THz generation from α-In$_2$Se$_3$ at 45° incident angle ($\theta$). The $\theta$ is described as the angle between the incident light and the normal direction of the specimen. The modulation of the light polarization can be changed by the half waveplate or quarter-wave plate (details in Supplementary Note 2). The $E_{xz}(t)$ (yellow curve) and $E_y(t)$ (purple curve) components of the emitted THz pulses are measured by a pair of wire-grid polarizers, which are expected to reconstruct an elliptically-polarized THz field $\boldsymbol{E}(t)$. Ultra-efficient THz signals are detected from a ferroelectric α-In$_2$Se$_3$ under the excitation of fs laser pulses, and the diameter of the pump laser beam on the sample is ~4 mm.

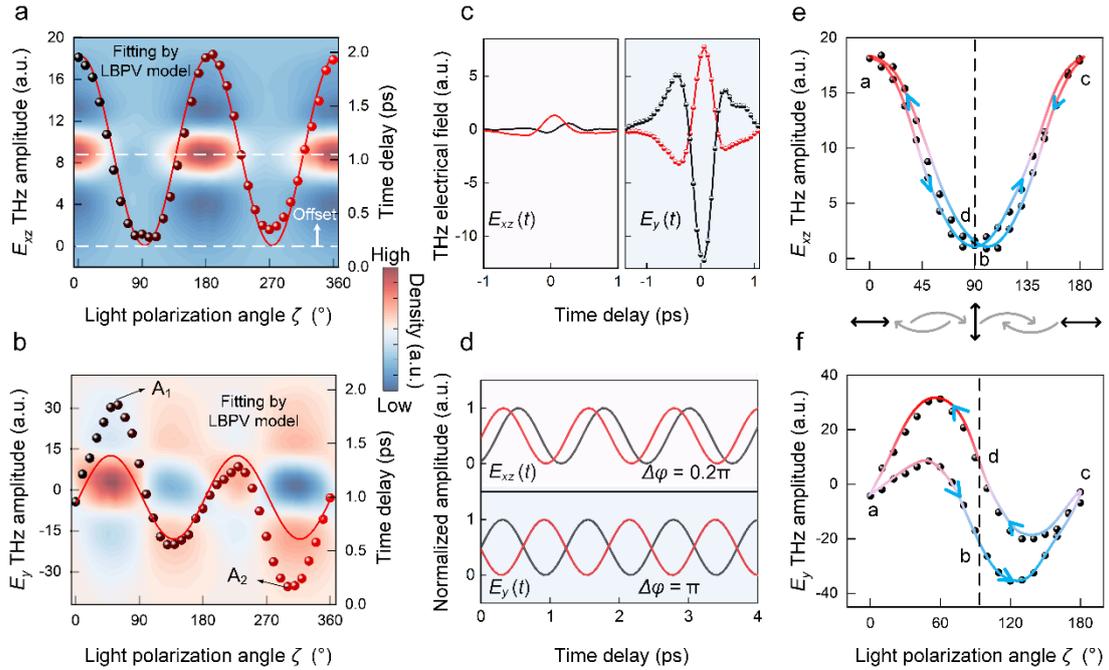

**Figure 2. Linearly-polarized-light induced the hysteresis of photocurrents in α-In₂Se₃.** The dependence of peak-valley values of the emitted (a) $E_{xz}(t)$ and (b) $E_y(t)$ THz signals on light polarization angle $\zeta$ at 45° oblique incidence. The red curve represents the theoretical results based on the LBPV tensorial analysis. (c) The comparison of $E_{xz}(t)$ (pink area) and $E_y(t)$ (blue area) THz electrical fields excited by 90° s-polarized light (red curve) and 270° s-polarized light (black curve). (d) The normalized electromagnetic wave at the center frequency of the THz wave in (c). The out-of-plane $E_{xz}(t)$ field has an obviously 0.2π-phase delay and in-plane $E_y(t)$ field presents π-phase reversal. (e-f) The $J_\lambda^{shift} - E_\xi$ ($\lambda=xz, y$) hysteresis loop (abcda) through changing the light

polarizations (0° →90°→180° →90° →0°).

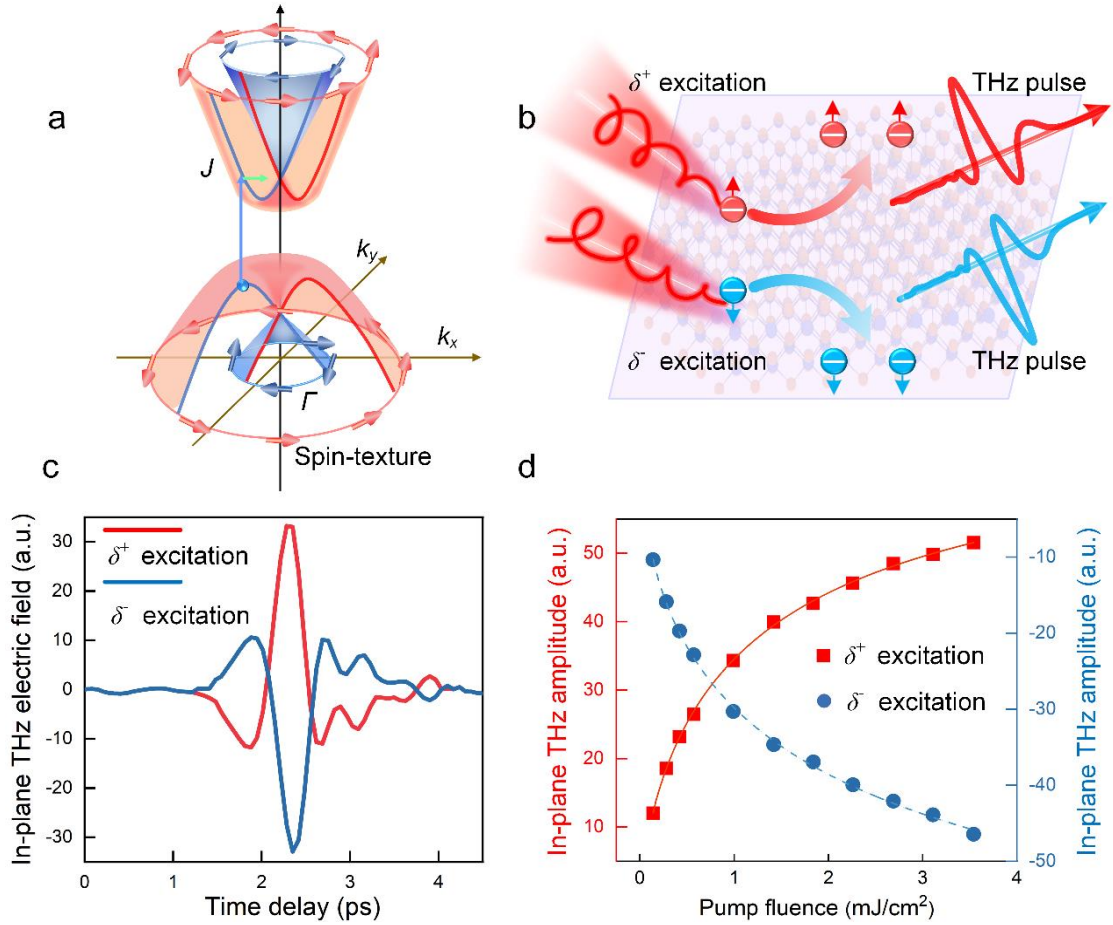

**Figure 3. Schematic diagrams of the CBPV effect based on Rashba spin-splitting and in-plane circular dichroism of ferroelectric α-In$_2$Se$_3$.** (a) The band diagram for spin-orientation-induced CBPV effect in α-In$_2$Se$_3$ with Rashba-type spin-splitting of energy band around $\Gamma$ point. Based on the optical selection rules, the circularly-polarized light (RCP ($\delta^+$) or LCP ($\delta^-$)) leads to nonequilibrium occupation of spin polarization, leading to a spin-polarized photocurrent $J$ (green line). Red and blue arrows represent momentum-dependent spin polarization with opposite directions. (b) Circularly-polarized light with a well-defined helicity will generate charges of spin-up ↑, while with opposite helicity of spin-down ↓, and finally causing opposite flowing path of photocurrents, thus generating THz wave with opposite polarity. (c) The time-domain waveform of in-plane component along y-axis ($E_y(t)$) generated from α-In$_2$Se$_3$ when illuminated with $\delta^+$ and $\delta^-$ light. (d) The dependence of the peak-valley values of $E_y(t)$ on the pump fluence of $\delta^+$ and $\delta^-$ light. The red and blue results in Fig. 3c,d represent the data under the $\delta^+$ and $\delta^-$ light excitation, respectively.

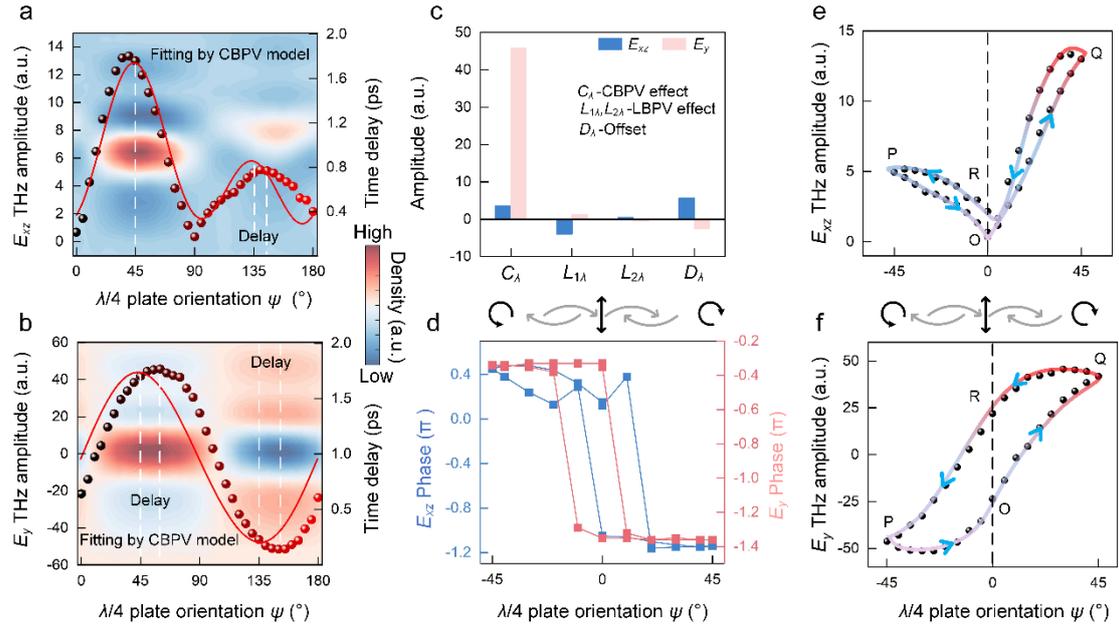

**Figure 4. Circularly-polarized-light induced the hysteresis of photocurrents from *α*-In$_2$Se$_3$.** The peak-valley values and the contour plots of (a) $E_{xz}(t)$ and (b) $E_y(t)$ measured for various setting of the λ/4 plate (QWP). The polarization angle *ψ* is the angle between the electrical field plane of the incident LP light and the fast axis of the QWP. The $ψ_{45°}$ and $ψ_{135°}$ represent the $δ^+$ and $δ^-$ light. The red curves represent the fitting results based on the CBPV theoretical prediction. (c) The extracted fitting coefficients $C_λ$, $L_{1λ}$, $L_{2λ}$, and $D_λ$ from the CBPV-based fitting results. The blue bar chart illustrates the four fitting coefficients of the $E_{xz}(t)$ component, while the red bar chart symbolizes the coefficients of $E_y(t)$. (d) The obtained polarization-dependent phase information of $E_{xz}(t)$ (blue curve) and $E_y(t)$ (red curve) through rotating the polarization angle *ψ* (0° →45°→0° →-45° →0°). The typical (e) $J_{xz}^C$ -$E_ψ$ and (f) $J_y^C$ -$E_ψ$ hysteresis loops (OQRPO) for a cyclical setting of *ψ*-values (0° →45°→0° →-45° →0°). Upon switching the light polarization, the helicity-dependent photocurrent does not return along its original path (POQ), but decreases along a curve slightly higher than the initial trajectory (QRP).

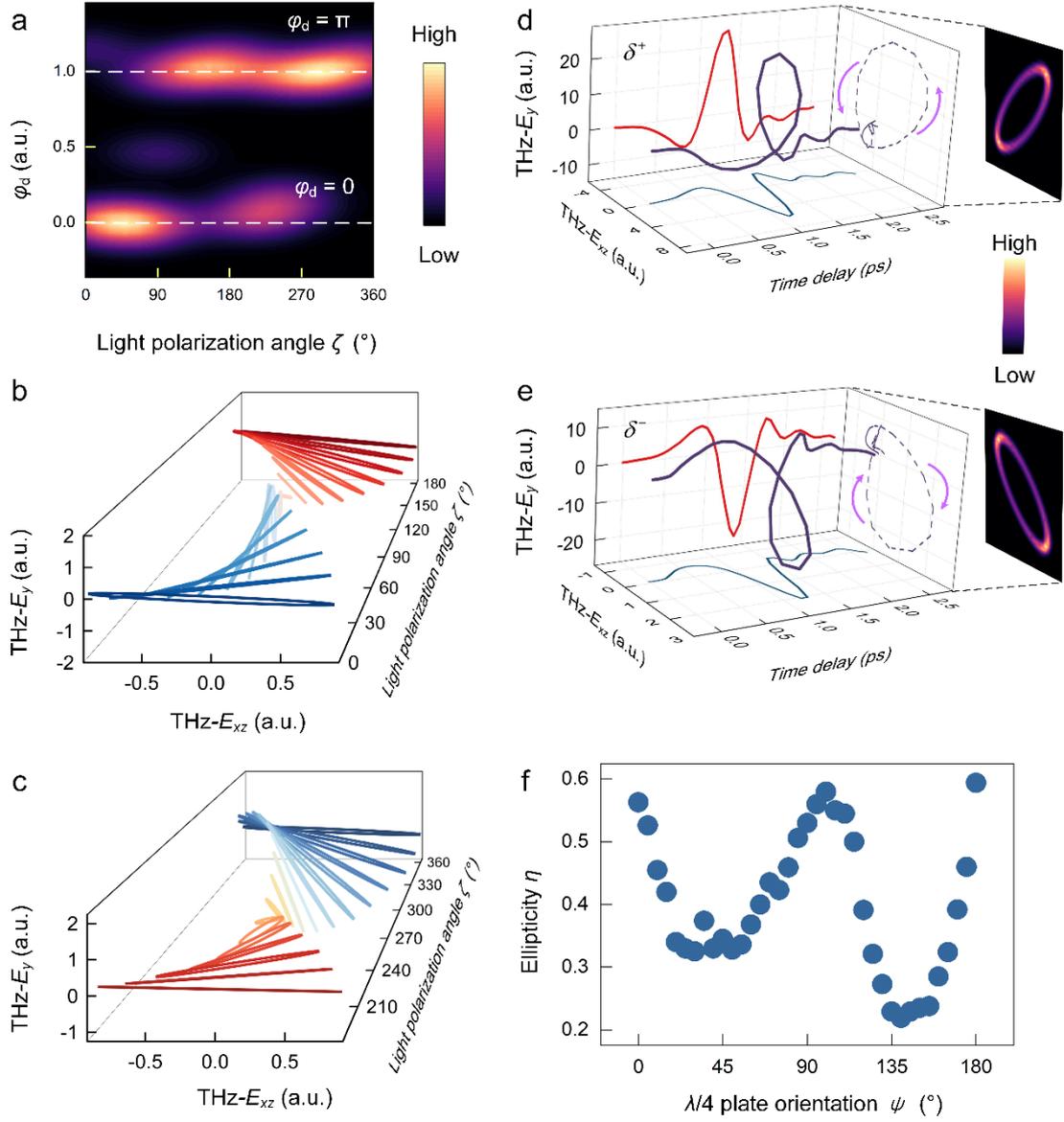

**Figure 5. Manipulation of elliptically-polarized THz electrical field by varying the light polarization.** (a) The dependence of relative phase ($\varphi_d$) between $E_{xz}(t)$ and $E_y(t)$ on the polarization angle $\zeta$ of linearly-polarized light. The polarization ellipse of reconstructed $\boldsymbol{E}(t)$ THz fields as $\zeta$ varies (b) from 0 to $\pi$ and (c) from $\pi$ to $2\pi$. Arbitrary manipulation of the chirality of $\boldsymbol{E}(t)$ from left-handed to right-handed is realized by changing the helicity of incident light. The 3D elliptical trajectory of THz field $\boldsymbol{E}(t)$ excited by (d) $\delta^+$ light and (e) $\delta^-$ light. The density mapping on the right side shows the $\boldsymbol{E}(t)$ polarization ellipse trajectory at the central frequency (1.2 THz), and its chirality is consistent with the time-domain 3D trajectory. (f) The ellipticity $\eta^{THz}$ of the polarization ellipse by varying the elliptically-polarized state of laser pulse through a QWP (from 0° to 180°). The experimental details of QWP manipulation can be found in Supplementary Note 9.

# Methods

**Quantum geometric photocurrents.**

In the modern theory of polarization, the nature of the macroscopic polarization is identified as a geometric phase (Berry phase) of the Bloch wavefunctions, revealing the topological nature of the ultrafast nonlinear photocurrent in ferroelectric materials lacking inversion symmetry[36]. The spontaneous electric polarization is evaluated as

$$P(k) = e \sum_{n \subset V} \int_{BZ} \frac{dk}{2\pi^3} a_n(k) \tag{1}$$

where $a_n(k) = -i\langle u_{nk} | \nabla_k u_{nk} \rangle$ (n: band index) is the intraband Berry connection of Bloch wave functions, and the integration over Brillouin zone (BZ) is known as a Berry phase. $u_{nk}(r)$ is the lattice periodic part of the Bloch wave $\psi_n = u_{nk}(r) e^{ik \cdot r}$. Photocurrents arise when the photoexcited electrons in a centrosymmetric-broken material under intense irradiation of monochromatic light. During the optical transition from valence band (VB: n=v) to conduction band (CB: n=c), the difference between $a_c(k)$ and $a_v(k)$ represents the real space shifting of photo-induced carriers[36]. This corresponds to the shift current $J_{shift} \propto \sigma^{(2)} E(\omega)^2$. The nonlinear photoconductivity $\sigma^{(2)}$ scales as the shift vector $R$ and interband transition dipole moment[37]

$$\sigma^{(2)}(0, \omega, -\omega) = \frac{2\pi e^3}{\hbar^2 \omega^2} \int_{BZ} \frac{d^3k}{(2\pi)^3} |\upsilon_{vc}|^2 R \delta(\omega_{cv} - \omega) \tag{2}$$

$$R = \text{Im}\left[\frac{(\partial_k \upsilon)_{vc}}{\upsilon_{vc}}\right] = \text{Im}[\partial_k (\log \upsilon_{vc})] + a_v - a_c \tag{3}$$

$\upsilon_{vc} = \langle u_v | \hat{\upsilon}_\alpha | u_c \rangle$ is the matrix element of the transition dipole moment from band $v$ to $c$; $\hbar\omega_{cv}=\varepsilon_c-\varepsilon_v$ represents the energy difference. The Eqs. 2-3 clearly indicate the quantum nature of shift current, which can be regarded as a visual probe of quantum mechanical phase in experiments.

The nonlinear optical response under the circularly-polarized light is well-known as injection current, which can be phenomenologically descript as: $J_q^C = i\gamma_{ql}(e \times e^*)_l I$. The subscript $q$ represents the photocurrent direction, $l$ symbolizes the light propagation direction. For circularly-polarized light propagating along $l$, $E = (E_r, e^{\pm i\pi/2}E_s, 0)$ with $E_r = E_s = E_0$ is the electric field vector. Then, $i(E \times E^*)_l = \pm(E_r E_s - E_s E_r)$, nonzero nonlinear response requires $E_r E_s \neq E_s E_r$. Thus, the intrinsic diagonal contribution for injection current would be nonvanishing. The circularly-polarized light induced photoconductivity of CBPV is[6]

$$\gamma_{q,rs}(0, \omega, -\omega) = -\frac{\pi e^3}{2\hbar^2} \int_{BZ} \frac{d^2k}{(2\pi)^2} \times \text{Im} \sum_{v,c} f_{vc} \Delta_{vc}^q [r_{cv}^r, r_{vc}^s] \delta(\omega_{cv} - \omega) \tag{4}$$

$f_{vc} = f_v - f_c$ and $\Delta_{vc}^q = v_{vv}^q - v_{cc}^q$ corresponds to occupation and velocity difference from band $v$ to $c$, respectively. $r_{cv}^s = \langle c | r^s | v \rangle$ descripts the interband transition dipole. $[r_{cv}^r, r_{vc}^s] = r_{cv}^r r_{vc}^s - r_{cv}^s r_{vc}^r = -i\varepsilon_{qrs}\Omega_{cv}^q$ represent interband Berry curvature ($\Omega$). Eq. 4 suggests that circularly-polarized light will pick up the interband Berry curvature ($\Omega$) between the valence and conduction band, and the injection current scales with velocity difference. The Berry curvature serves as imaginary part of quantum geometric tensor, namely, $T = g - i\Omega$. The Berry curvature is anti-symmetric in its tensor form. Hence, when the light helicity is reversed, the injection current flows oppositely with the magnitude unchanged (Fig. 3b). The CBPV could be utilized as a probing methodology to detect the ferroelectric Rashba spin-splitting.

**LBPV effect analysis based on the crystal symmetry.**

The detected orthogonal components of $E_{xz}(t)$ and $E_y(t)$ THz electric fields in our experimental setup can be expressed as

$$E_{xz}(t) \propto t_p \left( \frac{\partial J_x}{\partial t} \cos\theta_{THz} + \frac{\partial J_z}{\partial t} \sin\theta_{THz} \right), \quad E_y(t) \propto t_s \frac{\partial J_y}{\partial t} \tag{5}$$

where $\theta_{THz}$ is the refraction angle of THz wave in $\alpha$-In$_2$Se$_3$, satisfying general Snell's law: $n_1 \theta_{air} = n_2 \theta_{THz}$. $n_1$ and $n_2$ represent the refraction index of air and $\alpha$-In$_2$Se$_3$. $t_p$ and $t_s$ represent the Fresnel transmission coefficients at the In$_2$Se$_3$-air interface. The density of transient photocurrent $J(t)$ generated from BPV effect can be expressed as

$$J_q = J_q^L + J_q^C = \sigma_{qrs} E_r E_s^* + i\gamma_{ql}(e \times e^*)_l I \tag{6}$$

wherein $\sigma_{qrs}$ and $\gamma_{ql}$ represent the symmetric tensor and antisymmetric pseudo-tensor. $E_r$ and $E_s$ represent the electric field components of light. The $I = |E(q,\omega)|^2 cn_\omega / 2\pi$ is the pump light intensity and $\hat{e}$ is the unit polarization vector of the pump light. $J^L$ and $J^C$ represent the shift current and the injection current; and the corresponding BPV effect are dubbed as the LBPV effect (symmetric part) and CBPV effect (antisymmetric part), respectively. The $\alpha$-In$_2$Se$_3$ belongs to the $C_{3v}$ point group, exhibiting a typical non-centrosymmetric symmetry. When $\alpha$-In$_2$Se$_3$ is excited by the LP laser pulse ($P_C=0$), the $E_{xz}(t)$ and $E_y(t)$ THz fields arising from LBPV effect can be given as follows (details in Supplementary Note 4/Eqs. 7-16)

$$E_{xz}(t, \xi) \propto (\sigma_{xzx} \cos 2\xi + \sqrt{2}\sigma_{yxx} \sin 2\xi)\cos\theta_{THz} + \frac{1}{2}(\sigma_{zzz} \cos 2\xi - \sigma_{zxx} \cos 2\xi)\sin\theta_{THz} + M_1 \tag{7}$$

$$E_y(t, \xi) \propto \frac{3}{2}\sigma_{yxx} \cos 2\xi + \sqrt{2}\sigma_{xzx} \sin 2\xi - M_2 \tag{8}$$

where $\theta_{THz}$ is the refraction angle of THz wave in $\alpha$-In$_2$Se$_3$, $\xi$ is the angle describing the rotation of the electric field plane of LP light around $z$-axis, and $\sigma_{qrs}$ are the second-order nonlinear tensors. The terms of $M_1(\sigma_{xzx}\cos\theta_{THz} + (\sigma_{zzz} + 3\sigma_{zxx})\sin\theta_{THz}/2 + M_0)$ and $M_2(\sigma_{yxx}/2)$ are the fitting constants independent of the polarization angle, which could lead to upward or downward shift of THz amplitude.

**CBPV effect analysis based on the crystal symmetry.**

To clarify the origin of elliptically-polarized THz signal from $\alpha$-In$_2$Se$_3$, it is necessary to further prove the underlying mechanisms of the helicity dependent photocurrents. The $\mp\pi/2$ phase difference occurs between the orthogonal components by rotating the QWP to $\pm 45°$, which ultimately gives rise to the generation of RCP and

LCP light (see Supplementary Note 9). The second-order nonlinear optical response resulting from the excitation of circularly-polarized light is known as the CBPV effect. The injection current induced by the CBPV effect depends on the helicity of the incident circularly-polarized light, which can be expressed as

$$J_q^C = \gamma_{ql} P_C \hat{e}_l I = i\gamma_{ql}(e \times e^*)_l I, \ P_C \hat{e}_l = i(e \times e^*)_l \quad (9)$$

The injection current $J^C$ is proportional to the ellipticity $P_C$ of the incident light. $P_C = 0$ corresponds to the linearly-polarized light, and $P_C = +1$ and $P_C = -1$ corresponds to the RCP and LCP light. The non-zero CBPV pseudo-tensor of $\alpha$-In$_2$Se$_3$ are $\gamma_{xy}$ and $\gamma_{yx}$, then the photocurrent density component caused by CBPV are expressed as

$$J_x^C = \gamma_{xy} P_C \hat{e}_y I, \ J_y^C = \gamma_{yx} P_C \hat{e}_x I \quad (10)$$

Here we use a QWP to manipulate the helicity of the pump light. The helicity of the circularly-polarized light follows $P_C = \sin 2\psi$, where $\psi$ is the angle between the electric field plane of the LP light and the fast axis of the $\lambda/4$ waveplate. When the helicity of circular light is reversed, i.e., $P_C \rightarrow -P_C$, the rotation direction of circular photocurrent is also reversed for $J^C \rightarrow -J^C$. Such typical behavior of the circular photocurrent plays a significant role in confirming the CBPV response in the material system with broken inversion symmetry.

Based on the $C_{3v}$ symmetry of bulk $\alpha$-In$_2$Se$_3$, the THz radiation generated by the LBPV and the CBPV effect can be described as follow (details in Supplementary Note 4/Eqs. 19-22)

$$E_\lambda(t,\psi) \propto C_\lambda(t)\sin 2\psi + L_{1\lambda}(t)\cos 4\psi + L_{2\lambda}(t)\sin 4\psi + D_\lambda(t) \quad (11)$$

where $\lambda = y$ or $xz$. $C_\lambda(t)$ describe the CBPV response or the injection current, which presents $2\psi$ rotational symmetry of circularly-polarized light. Both $L_{1\lambda}$ and $L_{2\lambda}$ represent the LBPV response or the shift current, which present $4\psi$ dependence on the incident polarization state. The term $L_{1\lambda}$ and $L_{2\lambda}$ are associated with the nonzero second-order nonlinear tensors $\sigma_{xzx}$, $\sigma_{yxx}$, $\sigma_{zxx}$, and $\sigma_{zzz}$. $D_\lambda$ denotes the polarization independent term.

**Electric bands with Rashba-type spin-splitting and out-of-plane circular dichroism of ferroelectric α-In$_2$Se$_3$.**

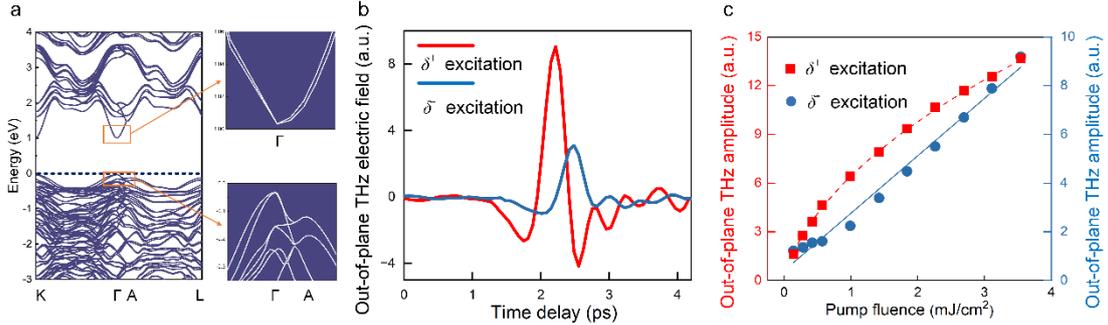

**Extended data Figure 1. Bulk bands with Rashba-type spin-splitting and out-of-plane circular dichroism of ferroelectric α-In$_2$Se$_3$.** (a) The electronic band structure of bulk $\alpha$-In$_2$Se$_3$ with spin orbit coupling through the Vienna Ab-initio Simulation Package. (b) Time-domain waveform of the out-of-plane ($E_{xz}$) THz pulse radiated from the $\alpha$-In$_2$Se$_3$ for RCP light ($\delta^+$) and LCP light ($\delta^-$) excitation. The pump fluence dependence of the amplitude of the out-of-plane THz electric field.

# Acknowledgement

This work was supported by National Natural Science Foundation of China (Nos. 12261141662, 12074311).